\documentclass[pre,twocolumn,amsmath,amssymb,superscriptaddress,floatfix]{revtex4}

\usepackage{graphicx}
\usepackage{dcolumn}
\usepackage{bm}

\begin{document}

\bibliographystyle{apsrev}

\title{Capillary bridging and long-range attractive forces 
       in a mean-field approach}

\author{D.~Andrienko}
\affiliation{Max-Planck-Institut f\"{u}r Polymerforschung, 
   Ackermannweg 10, D-55128 Mainz, Germany}

\author{P.~Patr{\'\i}cio}
\affiliation{Centro de F\'\i sica Te\'orica e Computacional, Universidade de
   Lisboa, Avenida Professor Gama Pinto 2, P-1649-003 Lisboa Codex,
   Portugal}
\affiliation{Instituto Superior de Engenharia de Lisboa, Rua
   Conselheiro Em\'\i dio Navarro 1, P-1949-014 Lisboa, Portugal}

\author{O.~I.~Vinogradova}
\affiliation{Max-Planck-Institut f\"{u}r Polymerforschung, 
   Ackermannweg 10, D-55128 Mainz, Germany}
\affiliation{Laboratory of Physical Chemistry of Modified Surfaces, 
   Institute of Physical Chemistry, Russian Academy of Sciences, 
   31 Leninsky Prospect, 119991 Moscow, Russia}

\date{\today}

\begin{abstract}
  When a mixture is confined, one of the phases can condense out.
  This condensate, which is otherwise metastable in the bulk, is
  stabilized by the presence of surfaces. In a sphere-plane geometry,
  routinely used in atomic force microscope (AFM) and surface force
  apparatus (SFA), it can form a bridge connecting the surfaces.  The
  pressure drop in the bridge gives rise to additional long-range
  attractive forces between them.
  Minimizing the free energy of a binary mixture we obtain the
  force-distance curves as well as the structural phase diagram of the
  configuration with the bridge. Numerical results predict a
  discontinuous transition between the states with and without the
  bridge and linear force-distance curves with hysteresis.
  We also show that similar phenomenon can be observed in a number of
  different systems, e.g. liquid crystals and polymer mixtures.
\end{abstract}

\pacs{}

\maketitle

\section{Introduction}

Capillary condensation is a classical example of a well-studied
phenomenon in nature~\cite{davis.ht:1996.a}. It is usually observed in
a porous medium with hydrophilic surfaces, where water condenses out
above its bulk transition temperature. The opposite effect, that is
capillary evaporation, is observed when the surfaces are hydrophobic.

First explained by Lord Kelvin for the case of vapor-water coexisting
phases, capillary condensation was also observed in other two phase
gas-liquid systems and studied in various confining
geometries~\cite{evans.r:1986.a,evans.r:1987.a,dobbs:1992.a,dobbs:1992.b}.
In fact, capillary condensation exists in any confined system close to
the transition point (binodal) provided that the surfaces prefer one
of the phases over the other and the bulk transition is of the first
order.  In this situation the condensed phase is stabilized by the
presence of the surfaces in the region of the phase diagram where the
condensed {\em bulk} state is metastable.

Interest in capillary condensation phenomenon has recently been
renewed due to the fast development of high-precision force measuring
devices, such as atomic force microscope (AFM) and surface force
apparatus (SFA). Indeed, accurate measurements of the interactions
between surfaces separated by thin fluid films led to a discovery of
unexpectedly strong and long-range attraction in some systems.  In
many cases, direct or indirect measurements proved that these
attractive interactions are due to the formation of a capillary
bridge. Below we mention some examples where, in our opinion, bridging
could play an important role:

(i) {\it attraction between hydrophobic surfaces:}  
\noindent
Early SFA measurements revealed the presence of an attractive
interaction between hydrophobic surfaces in aqueous
solutions~\cite{israelachvili:1982,rabinovich:1988,christenson:1988,parker:1994}.
These conclusions were confirmed by later AFM
experiments~\cite{meagher:1994,ishida:2000a,vinogradova:2001,attard:2003}.
This attraction is much larger than could be expected for a van der
Waals force.  As a typical example, the observed range of force for
silanated surfaces was up to $400\rm nm$ with the maximum attraction
or adhesion (normalized by radius of interacting surfaces) about $250
\rm mN/m$~\cite{ishida:2000a}.  It has soon became clear that
conventional theories of colloidal interactions fail to explain such a
long-range effect~\cite{eriksson:1989,podgornik:1989,attard:1989}.

Alternatively, it has been suggested that in some systems the
hydrophobic attraction is associated with the nucleation of dissolved
gas (gas-filled nanobubbles or drying film) at the
surfaces~\cite{vinogradova:1995}. When the surfaces become close,
these form a gaseous bridge between
them~\cite{carambassis:1998,horn:1999,yakubov:2000}, which generates
the attractive capillary force. This scenario was confirmed
indirectly, from the presence of the discontinuous steps in the force
curves, variability in the jump separations, hysteresis between
approach and retraction, and sensitivity to dissolved gas or another
solute~\cite{ishida:2000a,vinogradova:2001,butt:2003}, as well as from
more direct
experiments~\cite{bunkin:1997,gong:1999,tyrrell:2001,ishida:2002,steitz:2003}.

Contrary to the conclusions made on the basis of experimental data, no
consensus was achieved in the theoretical description of capillary
interactions.  Although on purely thermodynamic ground cavitation is
expected to occur between hydrophobic surfaces, both early
theory~\cite{yaminsky:1983} and later Monte Carlo
study~\cite{higashitani:2002} predict that this should happen only
when the separation becomes very small (i.e. comparable to the range
of van der Waals force).  The long-range capillary force was predicted
only for a situation, where pre-existing hemispherical bubbles (of
height and radius comparable to the range of expected attraction) are
postulated on an isolated surface~\cite{attard:2000}. Contrary to
this, it has been shown that the pre-formed nanobubbles are unstable
and tend to dissolve very
fast~\cite{eriksson:1995,ljunggren:1997,eriksson:1999}. This led to
speculations that the hydrophobic attraction is essentially a
non-equilibrium force~\cite{attard:2003}.

(ii) {\it nematic liquid crystals:} An exotic, but rather important
example of capillary bridging has recently been observed in nematic
liquid crystals~\cite{kosevar:2001.b,kosevar:2001.a,kosevar:2003.a}.
Here the two phases forming the bridge are both liquids with
approximately the same density; the main difference between them is a
degree of {\em orientational} molecular ordering: anisotropic
molecules are orientationally {\em disordered} in an isotropic phase;
in a nematic phase they are oriented along some preferred orientation,
called the director. The fact that the two phases have different
symmetry implies that the nematic-isotropic transition is of the
first-order~\cite{chaikin_lubensky}. It was
confirmed~\cite{stark.h:2004.a} that, in the frame of the Landau-de
Gennes mean-field approach, the bridge of the nematic phase is formed
between two spherical particles immersed in an isotropic phase, once
the system is close to the nematic-isotropic transition.
 
(iii) {\it various mixtures:} More complex in composition, strongly
confined fluids, when brought close to the phase separation, often
have similar features in the force-distance profiles: the force has an
unprecedentedly long-range attraction with a clear hysteresis on
retraction/approach. For instance, capillary bridging was observed in
bicontinuous microemulsions~\cite{petrov:1997} and in a solution of
polymer mixtures~\cite{wennerstrom:1998}.

The importance of capillary bridging as a possible driving force in
colloidal
systems~\cite{yushchenko:1983,dobbs:1992.a,dobbs:1992.b,bauer:2000}
cannot be overestimated.  The formation of the condensate between the
particles can trigger their aggregation and, as a result, flocculation
in colloidal suspensions. Small amount of added fluid generates,
through capillary effect, strong adhesive forces between the grains of
granular materials~\cite{JaegerNB96,HalseyL98,SchulzSH03}.

The direct optical measurements of the bridge parameters are
impossible, due to its submicroscopic size.  One can study only the
indirect consequences, e.g.  force-distance curves in AFM or SFA
experiments, and from there attempt to deduce some of the system
properties. This requires theoretical modeling that fills in the gap.

The aim of this paper is to study the general consequences of the
formation/annihilation of a capillary bridge upon change of the
thermodynamical parameters or geometry of the system. Considering that
recent experimental results on capillary condensation/evaporation are
performed using AFM or SFA setups, we work in a sphere-plane geometry
which is relevant for the above-mentioned techniques. We first treat
the problem in a sharp-interface limit and derive simple analytical
expressions for the structural phase diagram and forces in
Sec.~\ref{sec:sharp_interface}. Then, in Sec.~\ref{sec:phase_field},
we use the Landau mean-field approach and solve the full nonlinear
equations for the bridge shape numerically. In the final
Section~\ref{sec:liquid_crystals}, using simple mapping of the free
energies, we discuss implication of our results to nematic liquid
crystals.

\section{Sharp interface limit}
\label{sec:sharp_interface}
To begin with, we construct a crude analytical solution that will give
us some qualitative picture of what is happening with the bridge
when the geometry or thermodynamical parameters of the system change.
The approach we shall follow was proposed by
Petrov~et.~al.~\cite{petrov:1997}. To keep the description as simple
as possible, we neglect the
internal structure of the interfaces. Then our system involves domains
of well-defined phases separated by infinitely thin interfaces, so
called ``sharp-interface limit''~\cite{elder.kr:2001.a}.

\begin{figure}
\begin{center}
\includegraphics[width=5cm, angle=0]{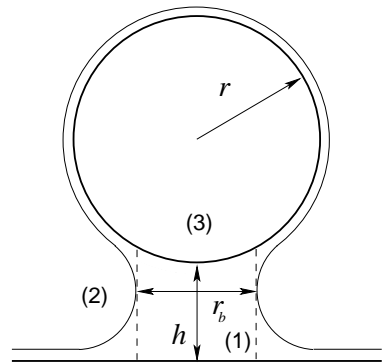}
\end{center}
\caption[]{
\label{fig:geometry} 
Illustration of the studied geometry and bridge parametrization.
Condensate of the phase (1) is stabilized between the apex of the sphere
and the flat substrate. Axial symmetry is assumed throughout the paper.}
\end{figure}

A geometry typical for SFA and AFM measurements is depicted in
Fig~\ref{fig:geometry}. At the moment of capillary condensation the
condensed phase is formed between the apex of the sphere and the flat
substrate. For convenience, we denote the phase inside the bridge as
a phase (1) and the bulk phase as a phase (2).  We also assume that the
sphere and the plane are made of the same material and denote it as
a phase (3). Then the excess free energy, i.e. the change in the free
energy of the system due to the emergence of the phase (1) (bridge) out of
the bulk phase can be written as the sum of the bulk and surface terms
\begin{equation}
\Delta G = G_b + G_s.
\label{eq:free_en}
\end{equation}
In the sharp-interface limit, the surface free energy $G_s$ is due to
different surface tensions of the interfaces between the phases 1/3
and 2/3, plus the energy of the interface between the phases (1) and
(2), which is proportional to the perimeter of the bridge
\begin{equation}
G_s = S_1(\sigma_{1,3}-\sigma_{2,3}) + S_2 \sigma_{1,2},
\end{equation}
where $S_1$ and $S_2$ are the areas of the interfaces between the
phases 1/3 and 1/2 correspondingly.

Additionally, we assume that the phase (1) wets the surfaces, i.e.
the sphere and the plane are covered by thin wetting layers of the
phase (1). This implies that the phase (1) has zero contact angle and
therefore, according to the Young equation
\begin{equation}
\sigma_{1,3}-\sigma_{2,3} = - \sigma_{1,2}.
\end{equation}
Note that here we implicitly assumed that we are below the prewetting
transition line, i.e. the layers of the metastable phase are thin and
we can neglect their contribution to the total free energy.

The bulk free energy is due to the difference in the chemical
potentials of the phase-separated component inside the bridge and in
the bulk
\begin{equation}
G_b = V \Delta \mu,
\label{eq:sharp_bulk}
\end{equation}
where $V$ is the volume of the bridge.

To calculate the surface areas and the volume of the bridge, its
circumference is approximated to a cylinder, instead of the correct,
concave shape. Similar approximation was used
before~\cite{attard:2000}. In fact, for the typical dimensions of the
surfaces and separations of SFA experiments this approximation works
very well~\cite{petrov:1997}.  The surface areas and the volume of the
bridge then read
\begin{eqnarray}
S_1 &=& 2 \pi r l + \pi r_b^2, \\
\nonumber
S_2 &=& 2 \pi r_b(l+h),       \\
\nonumber
V &=& \pi r_b^2 (l+h) - \frac{\pi}{6} l( 3 r_b^2 + l^2 ),
\end{eqnarray}
where $l = r - \sqrt{r^2 - r_b^2}$.

It is convenient to introduce a new constant with dimension of length
\begin{equation}
\lambda = \sigma_{1,2}/\Delta \mu,
\end{equation}
and scale out the variables with dimension of length by defining
\begin{eqnarray}
\label{eq:scaling}
\tilde{r} &=& r/\lambda, \\
\nonumber
\tilde{r_b} &=& r_b/\lambda, \\
\nonumber
\tilde{h} &=& h/\lambda,
\nonumber
\end{eqnarray}
We also introduce a dimensionless radial extent of the bridge which
plays a role of the order parameter in our system
\begin{equation}
x = r_b / r \in [0..1].
\end{equation}
With this notations, the excess free energy~(\ref{eq:free_en}) can be
written as
\begin{equation}
\Delta G(r, r_b, h, \lambda) = 
\Delta \mu \lambda^{3} G(x,\tilde{r},\tilde{h}). 
\end{equation}
For every fixed $\tilde{r}, \tilde{h}$, the minimum of
$G(x,\tilde{r},\tilde{h})$ gives the equilibrium radius of the bridge.

In what follows it will be understood that scaling~(\ref{eq:scaling})
has been carried out, and we shall omit the tildes in the text
below.

\begin{figure}
\begin{center}
\includegraphics[width=8cm, angle=0]{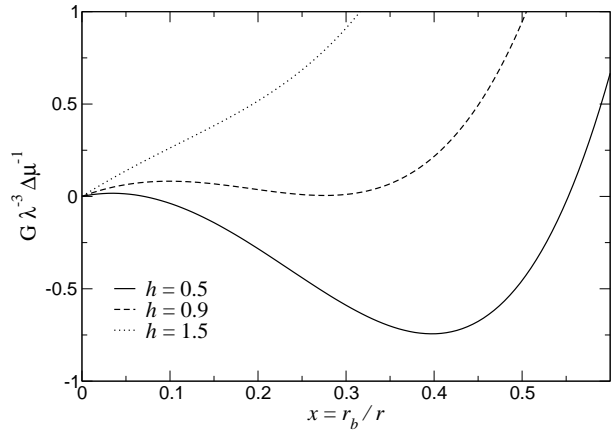}
\end{center}
\caption[]{
  Typical dependence of the excess free energy $G(x,r)$ on the radial
  extent of the bridge for $r/\lambda = 10$. Three separations are
  shown: $h=0.5 \lambda$, when the bridge is stable; $h=0.9\lambda$,
  with metastable bridge; $h=1.5\lambda$, when the configuration with
  the bridge is unstable.
\label{fig:sharp_energy} }
\end{figure}

Now that the mathematical problem is formulated, we first analyze the
dependence of the excess free energy on the bridge radius $r_b$, which
is shown in Fig.~\ref{fig:sharp_energy}. It is similar to the Landau
free energy as it occurs for phase transitions with a scalar order
parameter $x$. It predicts a discontinuous transition between the
bridged ($x > 0$) and unbridged ($x=0$) configurations on
approach/retraction of the sphere from the plane.  The first-order
character of the transition indicates possible jumps and hysteresis in
force-distance curves measured in AFM and SFA
experiments~\cite{ishida:2000a,vinogradova:2001,carambassis:1998,horn:1999,yakubov:2000}.

Let us now localize the transition line. The equilibrium radius of the
bridge corresponds to the minimum of the excess free energy
$G(x,r,h)$.  In addition, the configuration with the bridge is
energetically preferable if $G(x_t,r,h_t) < 0$.  Therefore, the
following equations determine the transition line (binodal in a
thermodynamical sense)
\begin{eqnarray}
\frac{\partial G}{\partial x}(x_t,r,h_t) = 0, G(x_t,r,h_t) = 0. 
\label{eq:binodal}
\end{eqnarray}

The configuration with the bridge (or without it) can also be {\em
  metastable}, when it is a {\em local} minimum of the excess free
energy. The corresponding metastability limits (spinodals) can be
determined from the condition
\begin{eqnarray}
\frac{\partial G}{\partial x}(x_m, r, h_m)  = 0,
\frac{\partial^2 G}{\partial x^2}(x_m, r, h_m) = 0.
\label{eq:spinodal}
\end{eqnarray}

\begin{figure}
\begin{center}
\includegraphics[width=8cm, angle=0]{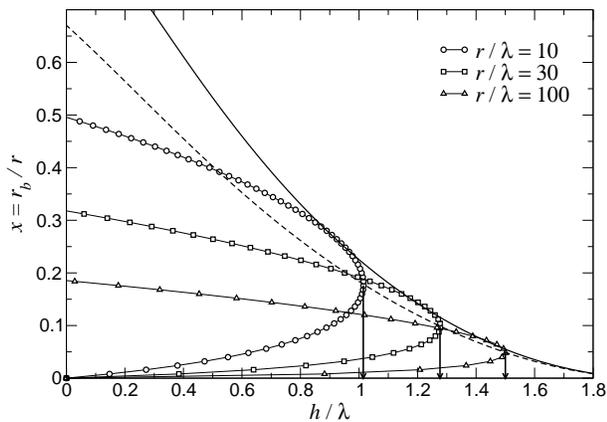}
\end{center}
\caption[]{
  Bridge radius $r_b / r$ as a function of the sphere-plane separation
  $h/\lambda$. The curves correspond to the different radii of the
  sphere $r/ \lambda = \{10, 30, 100\}$. Solid and dashed lines are
  binodal and spinodal correspondingly.
\label{fig:Rb_H}
}
\end{figure}

Now that we have a criteria for the stability and the metastability of
the bridge, Eqns.~(\ref{eq:binodal},\ref{eq:spinodal}), let us have a
look at the dependence of the bridge radius $r_b/r$ on the
sphere-plane separation, $h$.  If we differentiate the excess free
energy $G(x,r,h)$ with respect to $x$, we find that the extrema of the
free energy are solutions of the implicit equation for the bridge
radius
\begin{equation}
h = r\left( \sqrt{1-x^2}-1 \right) + 
\frac{x}{r^{-1}+x}\left( \sqrt{\frac{1-x}{1+x}} + 1 \right).
\label{eq:h_x}
\end{equation}
These solutions are shown in Fig.~\ref{fig:Rb_H} for several values of
the sphere radius $r$. As anticipated, there are two solutions for
each separation $h < h_m(r)$, which correspond to two extrema of the
excess free energy (see Fig.~\ref{fig:sharp_energy}). The larger of
these two values corresponds to the minimum of the free energy; its
value defines the radius of the stable (or metastable) bridge.  If the
separation $h$ is larger than some value, $h_m(r)$, then the solution
to Eq.~(\ref{eq:h_x}) does not exist.  Separation $h_m(r)$ provides,
therefore, metastability limit for the configuration with the bridge.

Calculating the first derivative of $G$ at $x = 0$ we find that
$\partial G / \partial x (x=0) = 2h \ge 0$, i.e. our model predicts
that the configuration without the bridge can be stable or metastable
and never unstable. Both the transition line,
Eq.~(\ref{eq:binodal}), and the metastability line,
Eq.~(\ref{eq:spinodal}), are shown in Fig.~\ref{fig:Rb_H}.

\begin{figure}
\begin{center}
\includegraphics[width=8cm, angle=0]{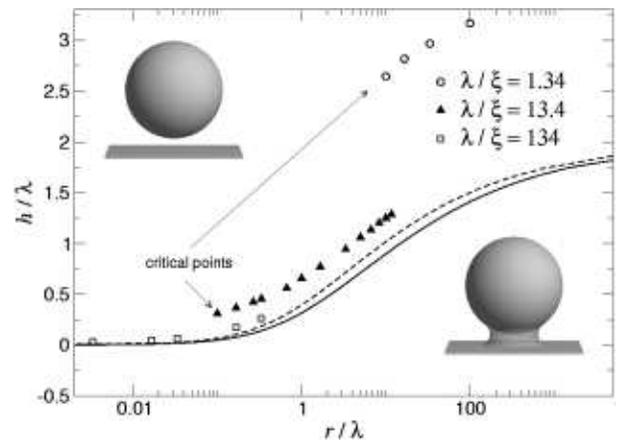}
\end{center}
\caption[]{
  Structural phase diagram of the capillary bridge.  For small $r /
  \lambda$ the bridge becomes unstable already for small separations
  $h/\lambda$.  For large $r/ \lambda$ the threshold separation
  asymptotically approaches $2 \lambda$.  Solid (dashed) line shows
  binodal (spinodal) correspondingly. Symbols presents the results of
  the numerical minimization. Note $\log$ scale for the sphere radius
  $r$.
\label{fig:diagram}
}
\end{figure}

Finally, we show the transition line (binodal) together with the
metastability limit (spinodal) on the structural phase diagram,
Fig.~\ref{fig:diagram}. It confirms that for small sphere radii the
configuration with the bridge becomes unstable almost immediately, for
very small sphere-plane separations. As the radius of the sphere
grows, the transition occurs at much larger separations, approaching
$h/\lambda = 2$ for $r \rightarrow \infty$. As a result, our model
predicts that bridging is not possible for $h > 2 \lambda$. The same
conclusion was made in Ref.~\cite{petrov:1997}.

After the structural phase diagram is calculated, let us have a look
at the force-distance curves.  The interaction force can be calculated
from the free energy 
\begin{equation}
\frac{ {\cal F} }{\Delta \mu \lambda^{2} } =
-\frac{\partial G(x,r,h)}{\partial h } = -\pi r (2 x+x^2 r),
\end{equation}
and, together with equation~(\ref{eq:h_x}), provides the
force-distance curves written in parametric form. 

\begin{figure}
\begin{center}
\includegraphics[width=8cm, angle=0]{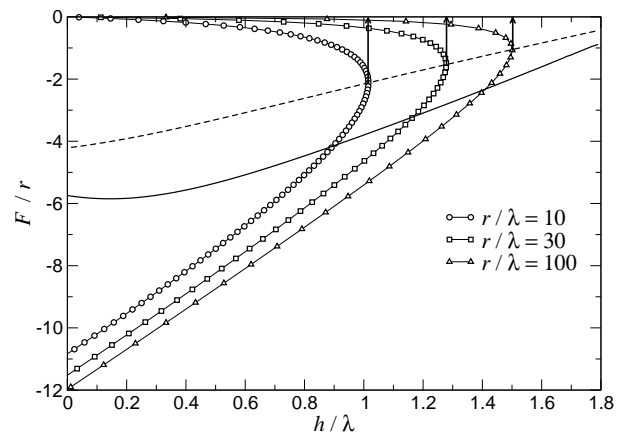}
\end{center}
\caption[]{
  Force-distance curves. Different symbols correspond to different
  radii of the sphere $r/ \lambda = \{10, 30, 100\}$. Solid and dashed
  lines are binodal and spinodals correspondingly.
\label{fig:sharp_force}
}
\end{figure}

This dependence is shown in Fig.~\ref{fig:sharp_force}, together with
metastability limits and the transition line. It quantifies our
previous conclusion that the transition is of the first order: for
small separations the configuration with the bridge is stable and
gives rise to practically linear increase of the force with
separation. At the transition point this configuration becomes
metastable and the bridge can disappear as soon as the fluctuations of
the bridge radius are big enough to destroy it. Finally, at even
larger separations, the bridge becomes completely unstable and
disappears. There is no interaction between the sphere and the plane
from this point on and the force jumps to the zero value. 

On the other hand, when the sphere approaches the surface, the
configuration without the bridge is stable (metastable) up to the zero
separation. Therefore, the largest hysteresis in force-distance curves
on approach/retraction is given by $h_m(r)$.

Before presenting the numerical results, let us briefly recall the
main conclusions. Assuming that the interface is infinitely sharp and
the bridge has a cylindrical shape, we are able to see the first-order
structural transition between the configurations with and without the
bridge. Structural phase diagram reveals that for small sphere radii
the transition occurs almost immediately, for small sphere-surface
separations. On the other hand, even for large sphere radii, bridging
is not possible for $h > 2 \sigma/\Delta \mu$. The first-order nature
of the transition leads to hysteresis in the the force-distance curves
on approach/retraction of the sphere from the surface. The attractive
force is basically proportional to the separation, with small
deviations close to the transition line.

\section{Mean-field approach}
\label{sec:phase_field}

It is clear that even the primitive model considered above allows a
qualitatively rich description of the transition.  However, there are
still several important details missing, on the first place: (i) the
non-cylindrical shape of the bridge; (ii) the finite thickness of the
interface; (iii) the interaction between the fluid and the solid walls.
Inclusion of these corrections makes the problem rather
involved and does not allow anymore for a simple analytical
solution. In what follows we use numerical methods to tackle the
problem.

\subsection{Free energy}

\label{sec:phase}

To describe the bulk phase as well as the interface structure,
mean-field theories~\cite{binder.k:1983.a} are often used.  In this
approach the order parameter $\phi$ is introduced. For a liquid-vapour
model $\phi$ is just the density. For a binary mixture $\phi$ is a
composition variable, defined as $\phi = (n_1 - n_2) / (n_1 + n_2)$,
where the $n_i$ are the number densities of the two species.  This
order parameter varies slowly in the bulk regions and rapidly on
length scales of the interfacial width. The unmixing thermodynamics is
described via a free energy functional.

In the mean-field approach the semi-grand potential of a binary mixture is
written as \cite{bonn.d:2001.a}
\begin{equation}
{\Omega}\{\phi\} =
\int d V \left[ \frac{k}{2} \left(\nabla \phi \right)^2
+ f(\phi) - \Delta \mu \phi \right],
\label{eq:grand_pot}
\end{equation}
where $f(\phi)$ is the Helmholtz free energy density of the
mixture, while $\Delta \mu$ is the chemical potential thermodynamically
conjugate to the order parameter $\phi$.

Since the material is confined in a container in any experiment, phase
separation is always affected by surface effects
\cite{evans.r:1986.a,evans.r:1987.a}. To include them, appropriate
surface terms responsible for the interaction of the liquid with the
container walls are added to the free energy
\cite{cahn.jw:1977.a,binder.k:1983.a,bonn.d:2001.a}.  We shall note
that the surface terms are very important: they give rise to a
prewetting transition and, together with $\Delta \mu$, determine the
thickness of the wetting layer. However, the prewetting transition
occurs only in the vicinity of the {\em bulk} phase separation
(binodal), i.e.  close to $\Delta \mu = 0$, while capillary bridging
can be observed in the whole region of the phase diagram between the
spinodal and the binodal. To avoid additional complications related to
the wetting/prewetting transitions, we assume that our system is below
the prewetting transition line (i.e.  we have thin wetting films) and
fix the surface value of the order parameter to its bulk value in the
infinite system.

The explicit form of the Helmholtz free energy $f(\phi)$ varies
depending on the type of mixture. However, the simple observation that
the two phases must coexist implies that there are two minima in the
free energy at the respective values of the order parameter. Here we
adopt the mean-field model for a symmetric mixture
\cite{reichl.le:1998.a,rowlinson.js:1969.a}
\begin{equation}
f(\phi) = - \frac{a}{2} \phi^2 + \frac{b}{4} \phi^4.
\label{eq:f_rm}
\end{equation}
Note that this is one of the simplest models to
describe unmixing; a more realistic description would need a more
sophisticated function, which also takes into account a dependence
on the overall density.

 The term $\left(\nabla \phi \right)^2$ is needed to provide spatial
structure to the theory: at phase coexistence, there are two bulk
equilibrium order parameter values $\phi_+$ and $\phi_-$ with the same
free energy density, $f(\phi_+) = f(\phi_-)$. Without the gradient
term, a structure with a large number of interfaces between the $\phi_+$ and
$\phi_-$ phase would be entropically favored. The term $(k/2)
\left(\nabla \phi \right)^2$ is the simplest one which penalizes
interfaces. While this is justified near the critical point, where
interfaces are very wide and the order parameter varies smoothly, a
more realistic description at strong segregation (where the interface
becomes rather sharp) would require higher--order gradients, too.

For the free energy~(\ref{eq:f_rm}) the binodal is given by $\Delta
\mu = 0$. Two spinodals (metastability limits) are located at $\Delta
\mu = \pm 2a^{3/2}/(27b)^{1/2}$.

Finally, we should take into account the interaction between
the solid walls and the fluid. To simplify our model,
we have chosen an interaction that acts locally at the solid-fluid interface,
imposing fixed boundary conditions at the walls, $\phi=\phi_-$.

\subsection{Minimization procedure}

The phase-field models are very convenient because no explicit
boundary tracking is needed.  However, they are wasteful in terms of
simulating bulk regions. One can remedy the situation by taking into
account that the order parameter varies slowly in bulk regions and
rapidly on length scales of the order of the correlation length, near
the interfaces. In this situation, finite elements method with
adaptive mesh size solves the problem of computational efficiency:
fine meshing is used only in the interfacial regions; bulk regions are
coarse-grained and have much larger size of finite elements.

The equilibrium distribution of the order parameter $\phi$ is obtained
by minimizing the free energy functional~(\ref{eq:grand_pot})
numerically using finite elements with adaptive meshes. During the
minimization the square integration region $L \times L$ was
triangulated using the $\rm BL2D$ subroutine~\cite{george.pl:1998.a}.
The function $\phi$ is set at all vertices of the mesh and is linearly
interpolated within each triangle. The free energy is then minimized
using the conjugate gradients method~\cite{press.wh:1992.a} under the
constraints imposed by the boundary conditions.

A new adapted mesh is generated iteratively from the previous
minimization. The new local triangle sizes are calculated from the
variations of the free energy, in order to guarantee a constant
numerical weight for each minimization variable~\cite{patricio.p:2002.a}. 
The final meshes,
with a minimal length of $\sim 10^{-3}$, had around $10^4$
minimization variables.

In order to obtain both stable and metastable configurations, we used
different types of initial conditions including configurations with
the capillary bridge and without it. To select stable solutions, we
calculated the grand potential of the mixture, $\Omega$, for both
stable and metastable solutions and chose the solution with the lowest
grand potential. This allows the accurate determination of the phase
diagram.

\section{Numerical results}

\subsection{Free interface}
\label{sec:free_interface}
To test the minimization algorithm and to tune the minimal mesh size
we first consider a system without a spherical particle and a surface.
We impose the boundary conditions in such a way that a flat interface
sets in the system, located at $z=0$, perpendicular to the $z$ axis.

In this geometry the order parameter $\phi$ depends only on the $z$
coordinate and the interface profile is a solution to the following
Euler-Lagrange equation
\begin{equation}
-k\frac{\partial^2 \phi}{\partial z^2} + \frac{\partial f}{\partial \phi} = 0,
\label{eq:interface_bp}
\end{equation}
together with the boundary conditions $\phi(\pm \infty) = \pm \phi_b$,
where $\pm \phi_b$ are the homogeneous bulk solutions, given by
$\partial f / \partial \phi = 0$
\begin{equation}
\phi_{\pm} = \pm \phi_b = \pm \sqrt{a/b}.
\end{equation} 
The solution to the boundary problem (\ref{eq:interface_bp}) yields
\begin{equation}
\phi = \phi_b \tanh \left( z/\xi  \right),
\label{eq:interface_sol}
\end{equation}
where $\xi = \sqrt{2k/a}$ is the correlation length, which serves as a
measure of the thickness of the interface.

With the solution~(\ref{eq:interface_sol}) we go back into the free
energy~(\ref{eq:grand_pot}), integrate over the $z$ coordinate and
obtain the interfacial tension $\sigma$, defined as the excess free energy
of the interface
\begin{equation}
\sigma = k \int_{-\infty}^{\infty}
\left(\frac{\partial \phi}{\partial z}\right)^2 dz = 
\frac{2}{3} a \phi_b^2 \xi.
\end{equation}

\begin{figure}
\begin{center}
\includegraphics[width=8cm, angle=0]{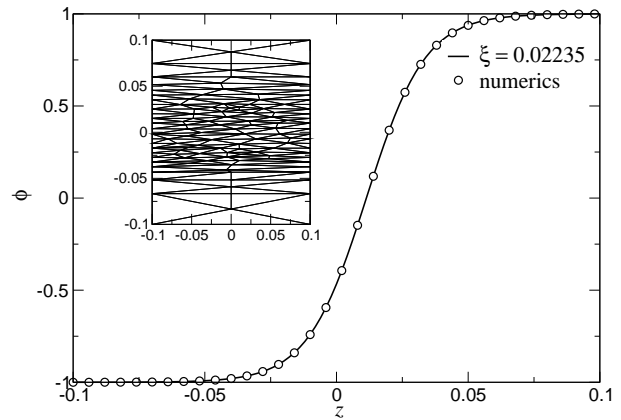}
\end{center}
\caption[]{
\label{fig:interface}
Order parameter profile of a free interface. Circles: results of the
numerical minimization; solid line: fit to
Eq.~(\ref{eq:interface_sol}) with $\xi = 0.02235$. Inset illustrates
the adaptive mesh: smaller triangles are used in the interface region
with strong variation of the order parameter. }
\end{figure}

For our numerical minimization we used the following parameters of the
potential~(\ref{eq:grand_pot}): $a = b = 4$, $k=0.001$. They yield the
correlation length $\xi \approx 0.02236$ and the interfacial tension
$\sigma \approx 0.06$.
 
The order parameter profile for this set of parameters is shown in
Fig.~\ref{fig:interface}. As expected, the order parameter changes
between two bulk values $\pm \phi_b = \pm 1$ and the width of the
interface is of the order of the correlation length $\xi$. A fit of
the results of the numerical minimization to
Eq.~(\ref{eq:interface_sol}) yields $\xi = 0.02235$, practically
indistinguishable from the exact value $\sqrt{2k/a}$.

The interface thickness $\xi$ sets the smallest length-scale in our
system and, therefore, defines the minimal size of the mesh. We found
that a minimal length of the final meshes ${\ell}= 10^{-3}$, which is
about 20 mesh points per interface, is accurate enough to recover the
interface structure.

The correlation length $\xi$ and the interfacial tension $\sigma$ are two
 parameters of our system which can be experimentally
measured. In addition, the interfacial tension enters the sharp-interface
description presented above. It is, therefore, convenient to rewrite
the excess free energy in terms of these two parameters. This shall
help us to compare the numerical results to the sharp-interface limit
as well as to the existing experiments.

Performing the substitution $\psi = \phi/\phi_b$ we obtain the
semi-grand potential density in the form
\begin{equation}
\omega  = \frac{3}{8}\frac{\sigma}{\xi} \left[
\xi^2 \left(\frac{\partial \psi}{\partial z} \right)^2 
- 2 \psi^2 + \psi^4 - 
\frac{4}{3} \frac{\xi}{\lambda} \psi \right].
\label{eq:grand_density}
\end{equation}
As before, we introduced a constant 
\begin{equation}
\lambda = \frac{\sigma }{2 \Delta \mu \phi_b}
\label{eq:phase_lambda}
\end{equation}
with dimension of length. Note that the length $\lambda$ has
(approximately) the same value as in the sharp-interface limit.  To
prove this, consider the situation when $\xi \to 0$. Then the
interface is sharp and the part of the excess free energy which scales
as the bridge volume can be written as 
\begin{equation}
f_{ex} = [ \Delta \mu (\phi_{+} -
\phi_{-}) + f( \phi_{-} ) - f( \phi_{+} ) ]V,
\end{equation}
where $V$ is the volume of the bridge. For small chemical potential
differences $\phi_{\pm} \approx \pm \phi_b$, where $\phi_b$ is the
value of the order parameter when $\Delta \mu = 0$. Then $f_{ex}$ can
be rewritten as $ 2 \Delta \mu \phi_b V$. Comparing $f_{ex}$ to the
expression for the bulk free energy in the sharp interface limit,
Eq.~(\ref{eq:sharp_bulk}), we obtain the definition of $\lambda$ in
the form of Eq.~(\ref{eq:phase_lambda}).

Now that the semi-grand potential density is written in a
dimensionless form it is clear that there is an additional
length-scale in the problem, when it is formulated in the phase-field
approach.  This length-scale is the correlation length $\xi$ or,
alternatively, the width of the interface between the two phases.

Three more variables with dimension of length are present in the
problem. Two of them, sphere radius $r$ and sphere-plane separation
$h$, fix the geometry of the system. For both SFA and AFM experiments
$r \gg h$. The third variable, $\lambda$, specifies the
thermodynamic state of the system.  The thickness of the interface
$\xi$ validates the sharp-interface limit: it is applicable only when
$\xi \ll \lambda, h, r$.

\subsection{Sphere-plane geometry}
We have thus seen that the interface thickness $\xi$ sets the
length-scale in our system. Therefore, three dimensionless ratios
uniquely specify the state of the system: $r / \xi$, $h / \xi$, and
$\lambda / \xi$. In what follows we fix the thickness of the interface
to $\xi= 0.02235$; the corresponding constants of the potential are
discussed in Section~\ref{sec:free_interface}. The rest of the
parameters with dimension of length (i.e. $r,h,\lambda$) are given in
units of $\xi$.

Let us first look at the situation when the thermodynamic
parameters of the system (i.e. $\lambda$, in addition to $\xi$) are
fixed, but the geometry is changing. For these calculations we used
the chemical potential difference $\Delta \mu = 0.1$, which yields
$\lambda / \xi \approx 13.4$.

The functional~(\ref{eq:grand_pot}) was minimized for a range of
sphere radii, $r / \xi \in [0.5..200]$ with the boundary conditions
$\phi=\phi_{-}$ at the sphere and the plane surfaces. Note that the
smallest size of the sphere is naturally set by the phenomenological
description we use: it is not applicable on length-scales much smaller
than the interface width $\xi$.  The upper value is limited by the
computational capacities: further increase in the sphere radius led to
significant slowdown of already rather time-consuming calculations.

\begin{figure}
\begin{center}
\includegraphics[width=8cm, angle=0]{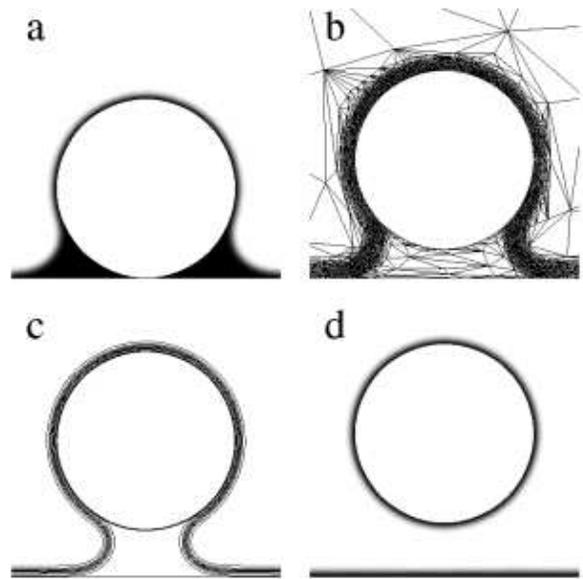}
\end{center}
\caption[]{
\label{fig:bridge} 
Cross-section of a capillary bridge for several surface-tip
separations: $h/\xi = \{ 0, 7.16, 11.63, 13.42 \}$. Sphere radius
$r/\xi=22.36$, $\lambda / \xi = 13.42$. b) illustrates the final mesh;
c) shows the interface contour lines, which are parallel to the
surface of the bridge.}
\end{figure}

Typical cross-sections of the bridge are shown in
Fig.~\ref{fig:bridge}. For small sphere-plane separations the wetting
film, covering the sphere and the plane, becomes thicker next to the
sphere, Fig.~\ref{fig:bridge}(a).  On retraction of the sphere from
the plane the bridge is formed and then gradually thinners, see
Figs.~\ref{fig:bridge}(b) and (c). Finally, at even larger
separations, the bridge becomes unstable and disappears. The
configuration with the sphere and the plane, wetted by a thin wetting
layer, Fig.~\ref{fig:bridge}(d), becomes energetically favorable.  At
these distances the sphere and the plane no longer interact with each
other.

\begin{figure}
\begin{center}
\includegraphics[width=8cm, angle=0]{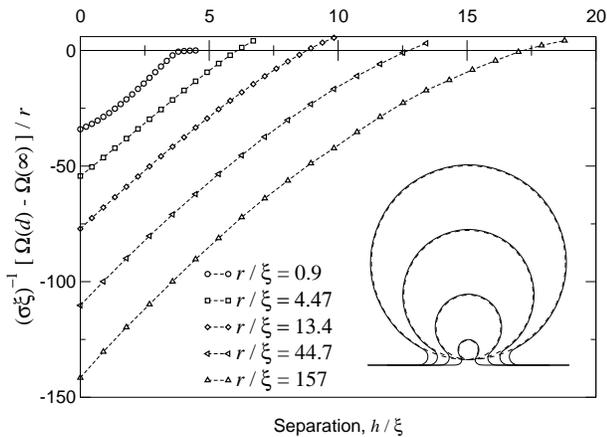}
\end{center}
\caption[]{
\label{fig:energy} 
Excess free energy of the configuration with the bridge vs sphere-plane
separation. Different curves correspond to different sphere radii.
Inset illustrates change in the radial extent and shape of the bridge
when the sphere radius changes. The inset configurations are taken at
a fixed sphere-surface separation, $h/\xi = 8.94$.}
\end{figure}

To investigate the nature of the structural transitions between the
bridged and unbridged configurations, we plot the excess free energy
as a function of the sphere-plane separation $h$ in
Fig.~\ref{fig:energy}.

The free energy profiles quantify the conclusions made above: when the
sphere is far from the plane, the configuration without the bridge is
the only stable one. The excess free energy does not change with
separation, i.e. there is no force acting on the sphere. When the
sphere moves closer to the plane, this configuration becomes first
metastable, and then unstable.  The bridge is formed and stays until
the sphere touches the surface.

On the other hand, on retraction, the configuration with the bridge
becomes first metastable and then unstable. Jump in the free energy
profile points on discontinuous (first-order) structural transition
between the configurations with and without the bridge. Due to the
first-order nature of the transition, the configuration with the
bridge can be kinetically stabilized and the bridge can, in principle,
disappear at any position where it is metastable. In fact, this
explains the variability in the jump-in/out distances observed in
experiment~\cite{carambassis:1998,ishida:2000a,yakubov:2000}.

On the face of it, the scenario predicted by the phase-field approach
is similar to the previous, sharp-interface limit one. However, in
addition to quantitative corrections, it also predicts {\em
  qualitatively} different results for small $r/\xi$. Indeed, as it is
seen from the free energy profile for $r/\xi = 0.9$, there is no jump
in the excess free energy. Instead, we have a smooth cross-over
between the bridged configuration and configuration without the
bridge. This points to a critical point $r_c$ below which the bridge
will no longer disappear discontinuously.  The difference between the
phase-field and the sharp-interface approaches for $r / \xi \approx 1$
is, of course, expected, since the infinitely sharp interface
assumption breaks down when the radius of the sphere is of the order
of the correlation length $\xi$. It is then tempting to explain the
absence of a jump in computer simulations~\cite{higashitani:2002},
where the system sizes are typically limited to thousands of angstroms.

To calculate the transition line, we fitted the free energy profiles
with a simple polynomial dependence (quadratic polynomial turned out
to be a reasonable approximation) and found where it intersects the
$\Omega=0$ axis. This allows accurate determination of the transition
line. The results are shown in Fig.~\ref{fig:diagram}, together with
the structural phase diagram obtained in the sharp-interface limit.
The agreement above the critical point is reasonable, taking into
account that we made rather crude approximations about the shape of the
bridge in the sharp interface limit.

\begin{figure}
\begin{center}
\includegraphics[width=8cm, angle=0]{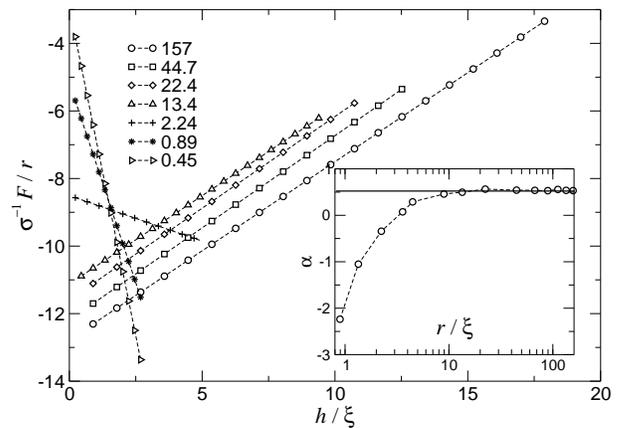}
\end{center}
\caption[]{
\label{fig:force} 
Force-distance curves calculated for several dimensionless sphere
radii $r/\xi$. Inset shows the slope of the curves $\alpha$ vs the
sphere radius $r$.}
\end{figure}

We have thus seen that the excess free energy profiles, scaled out by
the sphere radius, are accurately fitted by a quadratic
polynomial.  For large $r/\xi$ these profiles have {\em convex} shape
pointing that the attractive force, which is the first derivative of
the excess free energy with respect to the sphere-plane separation,
{\em decreases linearly} with the increase of the sphere-plane
separation. It can be also seen that the free energy profiles, when
scaled out by $r$, are almost parallel to each other, especially for
$r/\xi \to \infty$.  This means that the {\em slope} of the force,
scaled by the sphere radius $r$, approaches asymptotically a
constant value when $r/\xi \to \infty $. For smaller $r/\xi$ the
force profiles become first {\em linear} in $h$ and then start to curve
inwards, changing their shape to {\em concave}. As a result, the force
is independent of $h$ for some particular sphere radius; for even
smaller $r/\xi$ it {\em increases} with the increase of $h$.

This is illustrated in Fig.~\ref{fig:force}, where we show the
force-distance curves calculated for $\lambda / \xi = 13.4$.  The
inset shows the slope $\alpha$ obtained from the linear fit
\begin{equation}
 {\cal F} =  {\cal F}_0 + \sigma \xi \alpha r h.
\end{equation}
Note that here both $r$ and $h$ are dimensionless, i.~e. given in
units of $\xi$. The fit confirms that for $r / \xi > 10$ we have a
constant slope $\alpha \approx 0.5$, i.e. the force simply scales with
the sphere radius $r$, like in the Derjaguin
approximation~\cite{white:1983}.

We studied two more situations, with $\lambda / \xi = 134 \gg 1$ and
$\lambda / \xi = 1.34 \sim 1$.  The transition lines for both cases
are also shown on the structural phase diagram,
Fig.~\ref{fig:diagram}. For $\lambda / \xi = 134$ the agreement with
the sharp interface limit is remarkably better. At the same time, it
is clearly worse for $\lambda / \xi \sim 1$: the transition line is
shifted toward bigger separations and ends in a critical point for
rather large $r/\lambda$. There are two reasons for this. First, the
shape of the bridge is, in fact, a surface of revolution whose mean
curvature is constant. These surfaces are known as Delaunay surfaces,
or onduloids~\cite{cundy:1989}.  $\lambda$ defines the ``radius of
curvature'' of the bridge cross-section through its symmetry axis. If
$\lambda \gg r$ the bridge is well approximated by a cylinder. This
explains good agreement with the sharp interface limit for $\lambda /
\xi = 134$.  On the other hand, when $\lambda \sim \xi$ the radius of
curvature is of the order of the interface thickness, i.e. the sharp
interface limit cannot be used, even if $r,h \gg \xi$.

In spite of the complications in the structural phase diagram for
$\lambda \sim \xi$ and $r \sim \xi$, the main conclusions of our
numerical studies remain: the transition between bridged and unbridged
configurations is of the first order; the transition line ends in a
critical point located at $r_c/\xi \sim 1$. For large spheres, $r/ \xi
> 10$, the force is a linear function of the sphere-plane separation
$h$ with a slope proportional to the sphere radius $r$.

\section{Examples and discussion}
\label{sec:liquid_crystals}

In the previous sections a theoretical and numerical analysis of the
bridge formation and stability has been developed for a simple binary
mixture.  Here we discuss implication of our results for several
different systems.

We begin with nematic liquid crystals, mentioned in Introduction.
From the phenomenological point of view, the system is described by
the Landau-de Gennes free energy~\cite{degennes.pg:1995.a}
\begin{equation}
F\{\bm Q\} =
\int{(f_b + f_e)dV} + \int{f_s dS}
\label{eq:nem_free_en}
\end{equation}
where $f_b$ is the bulk free energy density, $f_e$ is the elastic free
energy density and $f_s$ is the surface free energy. Within a
mesoscopic approach the minimum of the Landau-de Gennes functional
$F\{\bm Q\}$ gives the equilibrium value of the tensor order parameter
$\bm Q$.

Symmetry arguments yield for the local bulk free energy
density~\cite{stephen.mj:1974.a,degennes.pg:1995.a}
\begin{equation}
f_b = a {\rm Tr} {\bm Q^2} -
      b {\rm Tr} {\bm Q^3} +
      c \left[{\rm Tr} {\bm Q^2}\right]^2,
\label{eq:f_b}
\end{equation}
where $a$ is assumed to depend linearly on the temperature, while the
positive constants $b,c$ are taken temperature independent.

It is convenient to scale out the variables by defining
\begin{eqnarray}
\tilde{Q}_{ij} = 6c/b Q_{ij}, \\ \nonumber
\tilde{f}_b = 24^2c^3/b^4 f_b.
\end{eqnarray}
It will be understood that such scaling has been carried out, and we
shall omit the overbars in the text below.

We also introduce a dimensionless temperature $\tau$ by defining
\begin{equation}
a = \tau b^2/24c.
\end{equation}

The elastic free energy density can be written as~\cite{stephen.mj:1974.a}
\begin{equation}
f_e = \frac{1}{2} L_1
\frac{\partial Q_{ij}}{\partial x_k} \frac{\partial Q_{ij}}{\partial x_k} +
\frac{1}{2} L_2
 \frac{\partial Q_{ij}}{\partial x_j} \frac{\partial Q_{ik}}{\partial x_k},
\end{equation}
where the constants $L_1$ and $L_2$ are related to Frank-Oseen elastic
constants by $K_{11} = K_{33} = 9Q_b^2(L_1 + L_2/2)/2$ and $K_{22} =
9Q_b^2L_1/2$ and $Q_b$ is the bulk nematic order parameter. The
sign of $L_2$ defines the preferred orientation of the director at the
NI interface. $L_2>0$ ($L_2<0$) favors planar (perpendicular)
anchoring~\cite{degennes.pg:1971.a}.

Now that the full free energy is specified, we perform several
simplifications. To begin with we neglect the non-uniformity of the
director distribution inside the bridge. This assumption is
quite reasonable for the studied geometry: we assume homeotropic
(perpendicular) anchoring of the director at the surfaces and planar
(parallel) anchoring at the nematic-isotropic interface. Second, we
assume that the nematic phase is uniaxial and neglect the biaxiality
of the interface. The full problem has been considered recently by
Stark et.~al.~\cite{stark.h:2004.a}.

For a {\em uniform uniaxial} nematic ($Q_{11} = Q$, $Q_{22} = Q_{33} =
-1/2Q$) the free energy (\ref{eq:f_b}) takes the form
\begin{equation}
f_{b} = \tau Q^2 - 2Q^3 + Q^4.
\label{eq:nem_uni_f}
\end{equation}
Free energy (\ref{eq:nem_uni_f}) predicts that the nematic state is
stable when $\tau < \tau_{\rm NI} = 1$ with a degree of orientational
order given by
\begin{equation}
Q_b = \frac{3}{4}\left(1+\sqrt{1-\frac{8}{9}\tau} \right),
\end{equation}
The same expression provides us with the metastability limit of the
nematic phase on heating at $\tau^* = 9/8$.

The solution which corresponds to a free interface is then given by
the de~Gennes Ansatz~\cite{degennes.pg:1971.a},
\begin{equation}
Q = \frac{1}{2}\left( 1+\tanh\frac{z}{\xi} \right),
\label{eq:ord_interface}
\end{equation}
where $\xi$ is the nematic correlation length when the director is
parallel to the interface
\begin{equation}
\xi^2 = \frac{8 L_1 c}{b^2} \left( 6 + \frac{L_2}{L_1}\right).
\label{eq:ord_interface_theory}
\end{equation}

A linear transformation of the order parameter, $Q = (1+\phi ) /2$,
reduces the Ansatz~(\ref{eq:ord_interface}) to the familiar
form~(\ref{eq:interface_sol}). The same substitution transforms the free
energy of a uniaxial nematic to the free energy of a binary mixture
\begin{equation}
f_b = [\phi^4 -2(3 - 2 \tau)\phi^2 + 8(\tau-1)\phi]/16.
\end{equation}
Close to the nematic-isotropic transition, $\tau_{\rm NI} = 1$, the
$\tau$-dependence of the coefficient at $\phi^2$ is not important, and
the free energy has exactly the same form as (\ref{eq:phase_lambda})
provided that
\begin{equation}
\lambda = \frac{\xi}{6(\tau - 1)}.
\end{equation}
This is an interesting result: recalling that the nematic-isotropic
transition occurs at $\tau_{\rm NI} = 1$ and the superheating
temperature of the nematic phase $\tau^* = 9/8$ we see that $\lambda
\ge 4/3\xi $, i.e.  $\lambda$ spans the whole range of values, from
the interface thickness $\xi$ at the superheating temperature, to the
system sizes, set by the sphere radius $r$ and the sphere-plane
separation $h$, and diverges at the nematic-isotropic transition. This
implies that, depending on the temperature, one can have either
discontinuous (first-order) or continuous transition between bridged
and unbridged configurations. Moreover, by changing the temperature of
the liquid crystal it is possible to control the range and the
strength of the attractive force between the particles in colloidal
suspensions.

Another `liquid crystalline' example for which our results are
relevant is drag of colloidal particles by the nematic-isotropic
interface. Recent experiments~\cite{west.jl:2002.a} revealed that the
nematic-isotropic interface is able to drag colloidal particles and,
as a result, form spatial patterns in the liquid crystal cell.
Theoretical reexamination of the experimental
data~\cite{andrienko:2004.a} demonstrated that the force exerted on a
colloidal particle is proportional to the penetration depth, and the
slope of the force scales as the particle size $r$.  This points to
rather unexpected mapping: here the bridge connects the colloidal
particle and the interface itself, since the position of the latter is
fixed by a small temperature gradient present in the system. The
straightforward implication of our results, which is also confirmed
experimentally~\cite{west.jl:2002.a}, is that it is more difficult for
the interface to capture small particles. Indeed, the attractive force
from the side of the interface scales with radius. In addition, the
barrier becomes smaller, completely vanishing at the critical radius
$r_c$.

Finally, we would like to mention that the standard Flory-Huggins
expression for a free energy of an incompressible polymer mixture
~\cite{binder:1994,flebbe.t:1996.a} can be expanded in powers of the
polymer volume fraction $\phi$. Close to the critical point it has the
same form as (\ref{eq:f_rm}) with $a = -k_B T /N + \chi /2$, $b=k_B
T/(3N)$, where $\chi$ is the Flory-Huggins parameter, $N$ is the
effective chain length. This means that the whole picture of bridging,
described here, can be adapted for the case of incompressible polymer
mixtures. In fact, the capillary-induced phase separation in binary
polymer solutions has been experimentally
observed~\cite{wennerstrom:1998} and theoretically studied using
mean-field lattice modeling~\cite{olsson.m:2004}.

In many ways the interface acts like a membrane under tension. In our
particular situation the formation of a bridge between the surface and
a particle resembles the process of wrapping a colloidal particle by a
membrane~\cite{boulbitch.a:2002.a,dezerno.m:2003.a}. A detailed
comparison requires, however, further analysis.

As a final remark, we remind that, by using rigid boundary conditions,
we will not observe possible prewetting transitions which have been
studied in detail for simpler
geometries~\cite{binder.k:1983.a,evans.r:1986.a,dietrich.s:1988.a}.
These transitions normaly occur close to the bulk phase separation,
i.e. when $\Delta \mu \approx 0$, or $\lambda \gg 1$. The
corresponding part of the phase diagram ($r/\lambda, h/\lambda \ll 1$)
can, therefore, be affected by these effects.  
On the other hand, when $r \rightarrow \infty$, we recover the
capillary condensation in a slab geometry~\cite{evans.r:1986.a}.  In
fact, the critical point has similar to the capillary condensation
origin: it occurs when the interfacial width is comparable to the
sphere-plane separation, i.~e. to the width of the capillary.

\section{Conclusions}
In conclusion, we have presented a detailed study of interactions
between the sphere and the plane due to formation of the capillary
bridge between them. We have shown that the structural transition
between the configurations with and without the bridge is
discontinuous and ends in a critical point for small particle radii.
We have also demonstrated that the attractive force is long-ranged and
is basically proportional to the sphere-plane separation and possible
hysteresis on approach/separation. 

\acknowledgments 

The advice of R.~Evans is gratefully acknowledged. H.~Stark is thanked
for providing the preprint of his paper. We also thank M.~Deserno,
B.~D{\"unweg}, and M.~Tamashiro for useful discussions. D.A.
acknowledges the support of the Alexander von Humboldt foundation.


\begin{thebibliography}{66}
\expandafter\ifx\csname natexlab\endcsname\relax\def\natexlab#1{#1}\fi
\expandafter\ifx\csname bibnamefont\endcsname\relax
  \def\bibnamefont#1{#1}\fi
\expandafter\ifx\csname bibfnamefont\endcsname\relax
  \def\bibfnamefont#1{#1}\fi
\expandafter\ifx\csname citenamefont\endcsname\relax
  \def\citenamefont#1{#1}\fi
\expandafter\ifx\csname url\endcsname\relax
  \def\url#1{\texttt{#1}}\fi
\expandafter\ifx\csname urlprefix\endcsname\relax\def\urlprefix{URL }\fi
\providecommand{\bibinfo}[2]{#2}
\providecommand{\eprint}[2][]{\url{#2}}

\bibitem[{\citenamefont{Davis}(1996)}]{davis.ht:1996.a}
\bibinfo{author}{\bibfnamefont{H.~T.} \bibnamefont{Davis}},
  \emph{\bibinfo{title}{Statistical mechanics of phases, interfaces, and thin
  films}} (\bibinfo{publisher}{VCH Publishers}, \bibinfo{address}{New York},
  \bibinfo{year}{1996}).

\bibitem[{\citenamefont{Evans et~al.}(1986)\citenamefont{Evans, Marconi, and
  Tarazona}}]{evans.r:1986.a}
\bibinfo{author}{\bibfnamefont{R.}~\bibnamefont{Evans}},
  \bibinfo{author}{\bibfnamefont{U.~M.~B.} \bibnamefont{Marconi}},
  \bibnamefont{and} \bibinfo{author}{\bibfnamefont{P.}~\bibnamefont{Tarazona}},
  \bibinfo{journal}{J. Chem. Phys.} \textbf{\bibinfo{volume}{84}},
  \bibinfo{pages}{2376} (\bibinfo{year}{1986}).

\bibitem[{\citenamefont{Evans and Marconi}(1987)}]{evans.r:1987.a}
\bibinfo{author}{\bibfnamefont{R.}~\bibnamefont{Evans}} \bibnamefont{and}
  \bibinfo{author}{\bibfnamefont{U.~M.~B.} \bibnamefont{Marconi}},
  \bibinfo{journal}{J. Chem. Phys.} \textbf{\bibinfo{volume}{86}},
  \bibinfo{pages}{7138} (\bibinfo{year}{1987}).

\bibitem[{\citenamefont{Dobbs and Yeomans}(1992)}]{dobbs:1992.a}
\bibinfo{author}{\bibfnamefont{H.~T.} \bibnamefont{Dobbs}} \bibnamefont{and}
  \bibinfo{author}{\bibfnamefont{J.~M.} \bibnamefont{Yeomans}},
  \bibinfo{journal}{Journal Of Physics-Condensed Matter}
  \textbf{\bibinfo{volume}{4}}, \bibinfo{pages}{10133} (\bibinfo{year}{1992}).

\bibitem[{\citenamefont{Dobbs et~al.}(1992)\citenamefont{Dobbs, Darbellay, and
  Yeomans}}]{dobbs:1992.b}
\bibinfo{author}{\bibfnamefont{H.~T.} \bibnamefont{Dobbs}},
  \bibinfo{author}{\bibfnamefont{G.~A.} \bibnamefont{Darbellay}},
  \bibnamefont{and} \bibinfo{author}{\bibfnamefont{J.~M.}
  \bibnamefont{Yeomans}}, \bibinfo{journal}{Europhysics Letters}
  \textbf{\bibinfo{volume}{18}}, \bibinfo{pages}{439} (\bibinfo{year}{1992}).

\bibitem[{\citenamefont{Israelachvili and Pashley}(1982)}]{israelachvili:1982}
\bibinfo{author}{\bibfnamefont{J.}~\bibnamefont{Israelachvili}}
  \bibnamefont{and} \bibinfo{author}{\bibfnamefont{R.}~\bibnamefont{Pashley}},
  \bibinfo{journal}{Nature} \textbf{\bibinfo{volume}{300}},
  \bibinfo{pages}{341} (\bibinfo{year}{1982}).

\bibitem[{\citenamefont{Rabinovich and Derjaguin}(1988)}]{rabinovich:1988}
\bibinfo{author}{\bibfnamefont{Y.~I.} \bibnamefont{Rabinovich}}
  \bibnamefont{and} \bibinfo{author}{\bibfnamefont{B.~V.}
  \bibnamefont{Derjaguin}}, \bibinfo{journal}{Colloids And Surfaces}
  \textbf{\bibinfo{volume}{30}}, \bibinfo{pages}{243} (\bibinfo{year}{1988}).

\bibitem[{\citenamefont{Christenson and Claesson}(1988)}]{christenson:1988}
\bibinfo{author}{\bibfnamefont{H.~K.} \bibnamefont{Christenson}}
  \bibnamefont{and} \bibinfo{author}{\bibfnamefont{P.~M.}
  \bibnamefont{Claesson}}, \bibinfo{journal}{Science}
  \textbf{\bibinfo{volume}{239}}, \bibinfo{pages}{390} (\bibinfo{year}{1988}).

\bibitem[{\citenamefont{Parker et~al.}(1994)\citenamefont{Parker, Claesson, and
  Attard}}]{parker:1994}
\bibinfo{author}{\bibfnamefont{J.~L.} \bibnamefont{Parker}},
  \bibinfo{author}{\bibfnamefont{P.~M.} \bibnamefont{Claesson}},
  \bibnamefont{and} \bibinfo{author}{\bibfnamefont{P.}~\bibnamefont{Attard}},
  \bibinfo{journal}{Journal Of Physical Chemistry}
  \textbf{\bibinfo{volume}{98}}, \bibinfo{pages}{8468} (\bibinfo{year}{1994}).

\bibitem[{\citenamefont{Meagher and Craig}(1994)}]{meagher:1994}
\bibinfo{author}{\bibfnamefont{L.}~\bibnamefont{Meagher}} \bibnamefont{and}
  \bibinfo{author}{\bibfnamefont{V.~S.~J.} \bibnamefont{Craig}},
  \bibinfo{journal}{Langmuir} \textbf{\bibinfo{volume}{10}},
  \bibinfo{pages}{2736} (\bibinfo{year}{1994}).

\bibitem[{\citenamefont{Ishida et~al.}(2000)\citenamefont{Ishida, Sakamoto,
  Miyahara, and Higashitani}}]{ishida:2000a}
\bibinfo{author}{\bibfnamefont{N.}~\bibnamefont{Ishida}},
  \bibinfo{author}{\bibfnamefont{M.}~\bibnamefont{Sakamoto}},
  \bibinfo{author}{\bibfnamefont{M.}~\bibnamefont{Miyahara}}, \bibnamefont{and}
  \bibinfo{author}{\bibfnamefont{K.}~\bibnamefont{Higashitani}},
  \bibinfo{journal}{Langmuir} \textbf{\bibinfo{volume}{16}},
  \bibinfo{pages}{5681} (\bibinfo{year}{2000}).

\bibitem[{\citenamefont{Vinogradova et~al.}(2001)\citenamefont{Vinogradova,
  Yakubov, and Butt}}]{vinogradova:2001}
\bibinfo{author}{\bibfnamefont{O.~I.} \bibnamefont{Vinogradova}},
  \bibinfo{author}{\bibfnamefont{G.~E.} \bibnamefont{Yakubov}},
  \bibnamefont{and} \bibinfo{author}{\bibfnamefont{H.~J.} \bibnamefont{Butt}},
  \bibinfo{journal}{Journal Of Chemical Physics}
  \textbf{\bibinfo{volume}{114}}, \bibinfo{pages}{8124} (\bibinfo{year}{2001}).

\bibitem[{\citenamefont{Attard}(2003)}]{attard:2003}
\bibinfo{author}{\bibfnamefont{P.}~\bibnamefont{Attard}},
  \bibinfo{journal}{Advances In Colloid And Interface Science}
  \textbf{\bibinfo{volume}{104}}, \bibinfo{pages}{75} (\bibinfo{year}{2003}).

\bibitem[{\citenamefont{Eriksson et~al.}(1989)\citenamefont{Eriksson,
  Ljunggren, and Claesson}}]{eriksson:1989}
\bibinfo{author}{\bibfnamefont{J.~C.} \bibnamefont{Eriksson}},
  \bibinfo{author}{\bibfnamefont{S.}~\bibnamefont{Ljunggren}},
  \bibnamefont{and} \bibinfo{author}{\bibfnamefont{P.~M.}
  \bibnamefont{Claesson}}, \bibinfo{journal}{Journal Of The Chemical
  Society-Faraday Transactions Ii} \textbf{\bibinfo{volume}{85}},
  \bibinfo{pages}{163} (\bibinfo{year}{1989}).

\bibitem[{\citenamefont{Podgornik}(1989)}]{podgornik:1989}
\bibinfo{author}{\bibfnamefont{R.}~\bibnamefont{Podgornik}},
  \bibinfo{journal}{Journal Of Chemical Physics} \textbf{\bibinfo{volume}{91}},
  \bibinfo{pages}{5840} (\bibinfo{year}{1989}).

\bibitem[{\citenamefont{Attard}(1989)}]{attard:1989}
\bibinfo{author}{\bibfnamefont{P.}~\bibnamefont{Attard}},
  \bibinfo{journal}{Journal Of Physical Chemistry}
  \textbf{\bibinfo{volume}{93}}, \bibinfo{pages}{6441} (\bibinfo{year}{1989}).

\bibitem[{\citenamefont{Vinogradova et~al.}(1995)\citenamefont{Vinogradova,
  Bunkin, Churaev, Kiseleva, Lobeyev, and Ninham}}]{vinogradova:1995}
\bibinfo{author}{\bibfnamefont{O.~I.} \bibnamefont{Vinogradova}},
  \bibinfo{author}{\bibfnamefont{N.~F.} \bibnamefont{Bunkin}},
  \bibinfo{author}{\bibfnamefont{N.~V.} \bibnamefont{Churaev}},
  \bibinfo{author}{\bibfnamefont{O.~A.} \bibnamefont{Kiseleva}},
  \bibinfo{author}{\bibfnamefont{A.~V.} \bibnamefont{Lobeyev}},
  \bibnamefont{and} \bibinfo{author}{\bibfnamefont{B.~W.}
  \bibnamefont{Ninham}}, \bibinfo{journal}{Journal Of Colloid And Interface
  Science} \textbf{\bibinfo{volume}{173}}, \bibinfo{pages}{443}
  (\bibinfo{year}{1995}).

\bibitem[{\citenamefont{Carambassis et~al.}(1998)\citenamefont{Carambassis,
  Jonker, Attard, and Rutland}}]{carambassis:1998}
\bibinfo{author}{\bibfnamefont{A.}~\bibnamefont{Carambassis}},
  \bibinfo{author}{\bibfnamefont{L.~C.} \bibnamefont{Jonker}},
  \bibinfo{author}{\bibfnamefont{P.}~\bibnamefont{Attard}}, \bibnamefont{and}
  \bibinfo{author}{\bibfnamefont{M.~W.} \bibnamefont{Rutland}},
  \bibinfo{journal}{Phys. Rev. Lett.} \textbf{\bibinfo{volume}{80}},
  \bibinfo{pages}{5357} (\bibinfo{year}{1998}).

\bibitem[{\citenamefont{Considine et~al.}(1999)\citenamefont{Considine, Hayes,
  and Horn}}]{horn:1999}
\bibinfo{author}{\bibfnamefont{R.~F.} \bibnamefont{Considine}},
  \bibinfo{author}{\bibfnamefont{R.~A.} \bibnamefont{Hayes}}, \bibnamefont{and}
  \bibinfo{author}{\bibfnamefont{R.~G.} \bibnamefont{Horn}},
  \bibinfo{journal}{Langmuir} \textbf{\bibinfo{volume}{15}},
  \bibinfo{pages}{1657} (\bibinfo{year}{1999}).

\bibitem[{\citenamefont{Yakubov et~al.}(2000)\citenamefont{Yakubov, Butt, and
  Vinogradova}}]{yakubov:2000}
\bibinfo{author}{\bibfnamefont{G.~E.} \bibnamefont{Yakubov}},
  \bibinfo{author}{\bibfnamefont{H.~J.} \bibnamefont{Butt}}, \bibnamefont{and}
  \bibinfo{author}{\bibfnamefont{O.~I.} \bibnamefont{Vinogradova}},
  \bibinfo{journal}{Journal Of Physical Chemistry B}
  \textbf{\bibinfo{volume}{104}}, \bibinfo{pages}{3407} (\bibinfo{year}{2000}).

\bibitem[{\citenamefont{Nguyen et~al.}(2003)\citenamefont{Nguyen, Nalaskowski,
  Miller, and Butt}}]{butt:2003}
\bibinfo{author}{\bibfnamefont{A.~V.} \bibnamefont{Nguyen}},
  \bibinfo{author}{\bibfnamefont{J.}~\bibnamefont{Nalaskowski}},
  \bibinfo{author}{\bibfnamefont{J.~D.} \bibnamefont{Miller}},
  \bibnamefont{and} \bibinfo{author}{\bibfnamefont{H.~J.} \bibnamefont{Butt}},
  \bibinfo{journal}{International Journal of Mineral Processing}
  \textbf{\bibinfo{volume}{72}}, \bibinfo{pages}{215} (\bibinfo{year}{2003}).

\bibitem[{\citenamefont{Bunkin et~al.}(1997)\citenamefont{Bunkin, Kiseleva,
  Lobeyev, Movchan, Ninham, and Vinogradova}}]{bunkin:1997}
\bibinfo{author}{\bibfnamefont{N.~F.} \bibnamefont{Bunkin}},
  \bibinfo{author}{\bibfnamefont{O.~A.} \bibnamefont{Kiseleva}},
  \bibinfo{author}{\bibfnamefont{A.~V.} \bibnamefont{Lobeyev}},
  \bibinfo{author}{\bibfnamefont{T.~G.} \bibnamefont{Movchan}},
  \bibinfo{author}{\bibfnamefont{B.~W.} \bibnamefont{Ninham}},
  \bibnamefont{and} \bibinfo{author}{\bibfnamefont{O.~I.}
  \bibnamefont{Vinogradova}}, \bibinfo{journal}{Langmuir}
  \textbf{\bibinfo{volume}{13}}, \bibinfo{pages}{3024} (\bibinfo{year}{1997}).

\bibitem[{\citenamefont{Gong et~al.}(1999)\citenamefont{Gong, Stearnes,
  Fornasiero, Hayes, and Ralston}}]{gong:1999}
\bibinfo{author}{\bibfnamefont{W.~Q.} \bibnamefont{Gong}},
  \bibinfo{author}{\bibfnamefont{J.}~\bibnamefont{Stearnes}},
  \bibinfo{author}{\bibfnamefont{D.}~\bibnamefont{Fornasiero}},
  \bibinfo{author}{\bibfnamefont{R.~A.} \bibnamefont{Hayes}}, \bibnamefont{and}
  \bibinfo{author}{\bibfnamefont{J.}~\bibnamefont{Ralston}},
  \bibinfo{journal}{Physical Chemistry Chemical Physics}
  \textbf{\bibinfo{volume}{1}}, \bibinfo{pages}{2799} (\bibinfo{year}{1999}).

\bibitem[{\citenamefont{Tyrrell and Attard}(2001)}]{tyrrell:2001}
\bibinfo{author}{\bibfnamefont{J.~W.~G.} \bibnamefont{Tyrrell}}
  \bibnamefont{and} \bibinfo{author}{\bibfnamefont{P.}~\bibnamefont{Attard}},
  \bibinfo{journal}{Phys. Rev. Lett.} \textbf{\bibinfo{volume}{8717}},
  \bibinfo{pages}{176104} (\bibinfo{year}{2001}).

\bibitem[{\citenamefont{Ishida et~al.}(2002)\citenamefont{Ishida, Sakamoto,
  Miyahara, and Higashitani}}]{ishida:2002}
\bibinfo{author}{\bibfnamefont{N.}~\bibnamefont{Ishida}},
  \bibinfo{author}{\bibfnamefont{M.}~\bibnamefont{Sakamoto}},
  \bibinfo{author}{\bibfnamefont{M.}~\bibnamefont{Miyahara}}, \bibnamefont{and}
  \bibinfo{author}{\bibfnamefont{K.}~\bibnamefont{Higashitani}},
  \bibinfo{journal}{Journal Of Colloid And Interface Science}
  \textbf{\bibinfo{volume}{253}}, \bibinfo{pages}{112} (\bibinfo{year}{2002}).

\bibitem[{\citenamefont{Steitz et~al.}(2003)\citenamefont{Steitz, Gutberlet,
  Hauss, Klosgen, Krastev, Schemmel, Simonsen, and Findenegg}}]{steitz:2003}
\bibinfo{author}{\bibfnamefont{R.}~\bibnamefont{Steitz}},
  \bibinfo{author}{\bibfnamefont{T.}~\bibnamefont{Gutberlet}},
  \bibinfo{author}{\bibfnamefont{T.}~\bibnamefont{Hauss}},
  \bibinfo{author}{\bibfnamefont{B.}~\bibnamefont{Klosgen}},
  \bibinfo{author}{\bibfnamefont{R.}~\bibnamefont{Krastev}},
  \bibinfo{author}{\bibfnamefont{S.}~\bibnamefont{Schemmel}},
  \bibinfo{author}{\bibfnamefont{A.~C.} \bibnamefont{Simonsen}},
  \bibnamefont{and} \bibinfo{author}{\bibfnamefont{G.~H.}
  \bibnamefont{Findenegg}}, \bibinfo{journal}{Langmuir}
  \textbf{\bibinfo{volume}{19}}, \bibinfo{pages}{2409} (\bibinfo{year}{2003}).

\bibitem[{\citenamefont{Yaminsky et~al.}(1983)\citenamefont{Yaminsky,
  Yushchenko, Amelina, and Shchukin}}]{yaminsky:1983}
\bibinfo{author}{\bibfnamefont{V.~V.} \bibnamefont{Yaminsky}},
  \bibinfo{author}{\bibfnamefont{V.~S.} \bibnamefont{Yushchenko}},
  \bibinfo{author}{\bibfnamefont{E.~A.} \bibnamefont{Amelina}},
  \bibnamefont{and} \bibinfo{author}{\bibfnamefont{E.~D.}
  \bibnamefont{Shchukin}}, \bibinfo{journal}{Journal Of Colloid And Interface
  Science} \textbf{\bibinfo{volume}{96}}, \bibinfo{pages}{301}
  (\bibinfo{year}{1983}).

\bibitem[{\citenamefont{Shinto et~al.}(2002)\citenamefont{Shinto, Uranishi,
  Miyahara, and Higashitani}}]{higashitani:2002}
\bibinfo{author}{\bibfnamefont{H.}~\bibnamefont{Shinto}},
  \bibinfo{author}{\bibfnamefont{K.}~\bibnamefont{Uranishi}},
  \bibinfo{author}{\bibfnamefont{M.}~\bibnamefont{Miyahara}}, \bibnamefont{and}
  \bibinfo{author}{\bibfnamefont{K.}~\bibnamefont{Higashitani}},
  \bibinfo{journal}{Journal Of Chemical Physics}
  \textbf{\bibinfo{volume}{116}}, \bibinfo{pages}{9500} (\bibinfo{year}{2002}).

\bibitem[{\citenamefont{Attard}(2000)}]{attard:2000}
\bibinfo{author}{\bibfnamefont{P.}~\bibnamefont{Attard}},
  \bibinfo{journal}{Langmuir} \textbf{\bibinfo{volume}{16}},
  \bibinfo{pages}{4455} (\bibinfo{year}{2000}).

\bibitem[{\citenamefont{Eriksson and Ljunggren}(1995)}]{eriksson:1995}
\bibinfo{author}{\bibfnamefont{J.~C.} \bibnamefont{Eriksson}} \bibnamefont{and}
  \bibinfo{author}{\bibfnamefont{S.}~\bibnamefont{Ljunggren}},
  \bibinfo{journal}{Langmuir} \textbf{\bibinfo{volume}{11}},
  \bibinfo{pages}{2325} (\bibinfo{year}{1995}).

\bibitem[{\citenamefont{Ljunggren and Eriksson}(1997)}]{ljunggren:1997}
\bibinfo{author}{\bibfnamefont{S.}~\bibnamefont{Ljunggren}} \bibnamefont{and}
  \bibinfo{author}{\bibfnamefont{J.~C.} \bibnamefont{Eriksson}},
  \bibinfo{journal}{Colloids And Surfaces A-Physicochemical And Engineering
  Aspects} \textbf{\bibinfo{volume}{130}}, \bibinfo{pages}{151}
  (\bibinfo{year}{1997}).

\bibitem[{\citenamefont{Eriksson and Ljunggren}(1999)}]{eriksson:1999}
\bibinfo{author}{\bibfnamefont{J.~C.} \bibnamefont{Eriksson}} \bibnamefont{and}
  \bibinfo{author}{\bibfnamefont{S.}~\bibnamefont{Ljunggren}},
  \bibinfo{journal}{Colloids And Surfaces A-Physicochemical And Engineering
  Aspects} \textbf{\bibinfo{volume}{159}}, \bibinfo{pages}{159}
  (\bibinfo{year}{1999}).

\bibitem[{\citenamefont{Kocevar et~al.}(2001)\citenamefont{Kocevar, Borstnik,
  Musevic, and Zumer}}]{kosevar:2001.b}
\bibinfo{author}{\bibfnamefont{K.}~\bibnamefont{Kocevar}},
  \bibinfo{author}{\bibfnamefont{A.}~\bibnamefont{Borstnik}},
  \bibinfo{author}{\bibfnamefont{I.}~\bibnamefont{Musevic}}, \bibnamefont{and}
  \bibinfo{author}{\bibfnamefont{S.}~\bibnamefont{Zumer}},
  \bibinfo{journal}{Physical Review Letters} \textbf{\bibinfo{volume}{86}},
  \bibinfo{pages}{5914} (\bibinfo{year}{2001}).

\bibitem[{\citenamefont{Kocevar and Musevic}(2001)}]{kosevar:2001.a}
\bibinfo{author}{\bibfnamefont{K.}~\bibnamefont{Kocevar}} \bibnamefont{and}
  \bibinfo{author}{\bibfnamefont{I.}~\bibnamefont{Musevic}},
  \bibinfo{journal}{Physical Review E} \textbf{\bibinfo{volume}{64}},
  \bibinfo{pages}{051711} (\bibinfo{year}{2001}).

\bibitem[{\citenamefont{Kocevar and Musevic}(2003)}]{kosevar:2003.a}
\bibinfo{author}{\bibfnamefont{K.}~\bibnamefont{Kocevar}} \bibnamefont{and}
  \bibinfo{author}{\bibfnamefont{I.}~\bibnamefont{Musevic}},
  \bibinfo{journal}{Chemphyschem} \textbf{\bibinfo{volume}{4}},
  \bibinfo{pages}{1049} (\bibinfo{year}{2003}).

\bibitem[{\citenamefont{Chaikin and Lubensky}(1995)}]{chaikin_lubensky}
\bibinfo{author}{\bibfnamefont{P.~M.} \bibnamefont{Chaikin}} \bibnamefont{and}
  \bibinfo{author}{\bibfnamefont{T.~C.} \bibnamefont{Lubensky}},
  \emph{\bibinfo{title}{Principles of Condensed Matter Physics}}
  (\bibinfo{publisher}{Cambridge University Press},
  \bibinfo{address}{Cambridge}, \bibinfo{year}{1995}).

\bibitem[{\citenamefont{Stark et~al.}(2004)\citenamefont{Stark, Fukuda, and
  Yokoyama}}]{stark.h:2004.a}
\bibinfo{author}{\bibfnamefont{H.}~\bibnamefont{Stark}},
  \bibinfo{author}{\bibfnamefont{J.}~\bibnamefont{Fukuda}}, \bibnamefont{and}
  \bibinfo{author}{\bibfnamefont{H.}~\bibnamefont{Yokoyama}},
  \bibinfo{journal}{Phys. Rev. Lett.}  (\bibinfo{year}{2004}),
  \bibinfo{note}{unpublished}.

\bibitem[{\citenamefont{Petrov et~al.}(1997)\citenamefont{Petrov, Olsson, and
  Wennerstrom}}]{petrov:1997}
\bibinfo{author}{\bibfnamefont{P.}~\bibnamefont{Petrov}},
  \bibinfo{author}{\bibfnamefont{U.}~\bibnamefont{Olsson}}, \bibnamefont{and}
  \bibinfo{author}{\bibfnamefont{H.}~\bibnamefont{Wennerstrom}},
  \bibinfo{journal}{Langmuir} \textbf{\bibinfo{volume}{13}},
  \bibinfo{pages}{3331} (\bibinfo{year}{1997}).

\bibitem[{\citenamefont{Wennerstr{\"o}m
  et~al.}(1998)\citenamefont{Wennerstr{\"o}m, Thuresson, Linse, and
  Freyssingeas}}]{wennerstrom:1998}
\bibinfo{author}{\bibfnamefont{H.}~\bibnamefont{Wennerstr{\"o}m}},
  \bibinfo{author}{\bibfnamefont{K.}~\bibnamefont{Thuresson}},
  \bibinfo{author}{\bibfnamefont{P.}~\bibnamefont{Linse}}, \bibnamefont{and}
  \bibinfo{author}{\bibfnamefont{E.}~\bibnamefont{Freyssingeas}},
  \bibinfo{journal}{Langmuir} \textbf{\bibinfo{volume}{14}},
  \bibinfo{pages}{5664} (\bibinfo{year}{1998}).

\bibitem[{\citenamefont{Yushchenko et~al.}(1983)\citenamefont{Yushchenko,
  Yaminsky, and Shchukin}}]{yushchenko:1983}
\bibinfo{author}{\bibfnamefont{V.~S.} \bibnamefont{Yushchenko}},
  \bibinfo{author}{\bibfnamefont{V.~V.} \bibnamefont{Yaminsky}},
  \bibnamefont{and} \bibinfo{author}{\bibfnamefont{E.~D.}
  \bibnamefont{Shchukin}}, \bibinfo{journal}{Journal Of Colloid And Interface
  Science} \textbf{\bibinfo{volume}{96}}, \bibinfo{pages}{307}
  (\bibinfo{year}{1983}).

\bibitem[{\citenamefont{Bauer et~al.}(2000)\citenamefont{Bauer, Bieker, and
  Dietrich}}]{bauer:2000}
\bibinfo{author}{\bibfnamefont{C.}~\bibnamefont{Bauer}},
  \bibinfo{author}{\bibfnamefont{T.}~\bibnamefont{Bieker}}, \bibnamefont{and}
  \bibinfo{author}{\bibfnamefont{S.}~\bibnamefont{Dietrich}},
  \bibinfo{journal}{Phys. Rev. E} \textbf{\bibinfo{volume}{62}},
  \bibinfo{pages}{5324} (\bibinfo{year}{2000}).

\bibitem[{\citenamefont{Jaeger et~al.}(1996)\citenamefont{Jaeger, Nagel, and
  Behringer}}]{JaegerNB96}
\bibinfo{author}{\bibfnamefont{H.~M.} \bibnamefont{Jaeger}},
  \bibinfo{author}{\bibfnamefont{S.~R.} \bibnamefont{Nagel}}, \bibnamefont{and}
  \bibinfo{author}{\bibfnamefont{R.~P.} \bibnamefont{Behringer}},
  \bibinfo{journal}{Reviews Of Modern Physics} \textbf{\bibinfo{volume}{68}},
  \bibinfo{pages}{1259} (\bibinfo{year}{1996}).

\bibitem[{\citenamefont{Halsey and Levine}(1998)}]{HalseyL98}
\bibinfo{author}{\bibfnamefont{T.~C.} \bibnamefont{Halsey}} \bibnamefont{and}
  \bibinfo{author}{\bibfnamefont{A.~J.} \bibnamefont{Levine}},
  \bibinfo{journal}{Physical Review Letters} \textbf{\bibinfo{volume}{80}},
  \bibinfo{pages}{3141} (\bibinfo{year}{1998}).

\bibitem[{\citenamefont{Schulz et~al.}(2003)\citenamefont{Schulz, Schulz, and
  Herminghaus}}]{SchulzSH03}
\bibinfo{author}{\bibfnamefont{M.}~\bibnamefont{Schulz}},
  \bibinfo{author}{\bibfnamefont{B.~M.} \bibnamefont{Schulz}},
  \bibnamefont{and}
  \bibinfo{author}{\bibfnamefont{S.}~\bibnamefont{Herminghaus}},
  \bibinfo{journal}{Physical Review E} \textbf{\bibinfo{volume}{67}}
  (\bibinfo{year}{2003}).

\bibitem[{\citenamefont{Elder et~al.}(2001)\citenamefont{Elder, Grant,
  Provatas, and Kosterlitz}}]{elder.kr:2001.a}
\bibinfo{author}{\bibfnamefont{K.~R.} \bibnamefont{Elder}},
  \bibinfo{author}{\bibfnamefont{M.}~\bibnamefont{Grant}},
  \bibinfo{author}{\bibfnamefont{N.}~\bibnamefont{Provatas}}, \bibnamefont{and}
  \bibinfo{author}{\bibfnamefont{J.~M.} \bibnamefont{Kosterlitz}},
  \bibinfo{journal}{Phys. Rev. E} \textbf{\bibinfo{volume}{6402}},
  \bibinfo{pages}{021604} (\bibinfo{year}{2001}).

\bibitem[{\citenamefont{Binder}(1983)}]{binder.k:1983.a}
\bibinfo{author}{\bibfnamefont{K.}~\bibnamefont{Binder}}, in
  \emph{\bibinfo{booktitle}{Phase Transitions and Critical Phenomena}}, edited
  by \bibinfo{editor}{\bibfnamefont{C.}~\bibnamefont{Domb}} \bibnamefont{and}
  \bibinfo{editor}{\bibfnamefont{J.}~\bibnamefont{Lebowitz}}
  (\bibinfo{publisher}{Academic Press}, \bibinfo{year}{1983}),
  vol.~\bibinfo{volume}{8}, pp. \bibinfo{pages}{2--144}.

\bibitem[{\citenamefont{Bonn and Ross}(2001)}]{bonn.d:2001.a}
\bibinfo{author}{\bibfnamefont{D.}~\bibnamefont{Bonn}} \bibnamefont{and}
  \bibinfo{author}{\bibfnamefont{D.}~\bibnamefont{Ross}},
  \bibinfo{journal}{Rep. Prog. Phys.} \textbf{\bibinfo{volume}{64}},
  \bibinfo{pages}{1085} (\bibinfo{year}{2001}).

\bibitem[{\citenamefont{Cahn}(1977)}]{cahn.jw:1977.a}
\bibinfo{author}{\bibfnamefont{J.~W.} \bibnamefont{Cahn}}, \bibinfo{journal}{J.
  Chem. Phys.} \textbf{\bibinfo{volume}{66}}, \bibinfo{pages}{3667}
  (\bibinfo{year}{1977}).

\bibitem[{\citenamefont{Reichl}(1998)}]{reichl.le:1998.a}
\bibinfo{author}{\bibfnamefont{L.~E.} \bibnamefont{Reichl}},
  \emph{\bibinfo{title}{A modern course in statistical physics}}
  (\bibinfo{publisher}{John Wiley \& Sons}, \bibinfo{address}{New York},
  \bibinfo{year}{1998}).

\bibitem[{\citenamefont{Rowlinson}(1969)}]{rowlinson.js:1969.a}
\bibinfo{author}{\bibfnamefont{J.~S.} \bibnamefont{Rowlinson}},
  \emph{\bibinfo{title}{Liquids and liquid mixtures}}
  (\bibinfo{publisher}{Butterworth}, \bibinfo{address}{London},
  \bibinfo{year}{1969}), \bibinfo{edition}{2nd} ed.

\bibitem[{\citenamefont{George and Borouchaki}(1998)}]{george.pl:1998.a}
\bibinfo{author}{\bibfnamefont{P.~L.} \bibnamefont{George}} \bibnamefont{and}
  \bibinfo{author}{\bibfnamefont{H.}~\bibnamefont{Borouchaki}},
  \emph{\bibinfo{title}{Delaunay triangulation and meshing : application to
  finite elements}} (\bibinfo{publisher}{Hermes}, \bibinfo{address}{Paris},
  \bibinfo{year}{1998}).

\bibitem[{\citenamefont{Press et~al.}(1992)\citenamefont{Press, Flannery,
  Teukolsky, and Vetterling}}]{press.wh:1992.a}
\bibinfo{author}{\bibfnamefont{W.~H.} \bibnamefont{Press}},
  \bibinfo{author}{\bibfnamefont{B.~P.} \bibnamefont{Flannery}},
  \bibinfo{author}{\bibfnamefont{S.~A.} \bibnamefont{Teukolsky}},
  \bibnamefont{and} \bibinfo{author}{\bibfnamefont{W.~T.}
  \bibnamefont{Vetterling}}, \emph{\bibinfo{title}{Numerical Recipes in
  Fortran}} (\bibinfo{publisher}{Cambridge University Press},
  \bibinfo{address}{Cambridge}, \bibinfo{year}{1992}), \bibinfo{edition}{2nd}
  ed.

\bibitem[{\citenamefont{Patr{\'\i}cio et~al.}(2002)\citenamefont{Patr{\'\i}cio,
  Tasinkevych, and Telo~da Gama}}]{patricio.p:2002.a}
\bibinfo{author}{\bibfnamefont{P.}~\bibnamefont{Patr{\'\i}cio}},
  \bibinfo{author}{\bibfnamefont{M.}~\bibnamefont{Tasinkevych}},
  \bibnamefont{and} \bibinfo{author}{\bibfnamefont{M.~M.} \bibnamefont{Telo~da
  Gama}}, \bibinfo{journal}{Eur. Phys. J. E} \textbf{\bibinfo{volume}{7}},
  \bibinfo{pages}{117} (\bibinfo{year}{2002}).

\bibitem[{\citenamefont{White}(1983)}]{white:1983}
\bibinfo{author}{\bibfnamefont{L.~R.} \bibnamefont{White}},
  \bibinfo{journal}{Journal Of Colloid And Interface Science}
  \textbf{\bibinfo{volume}{95}}, \bibinfo{pages}{286} (\bibinfo{year}{1983}).

\bibitem[{\citenamefont{Cundy and Rollett}(1989)}]{cundy:1989}
\bibinfo{author}{\bibfnamefont{H.}~\bibnamefont{Cundy}} \bibnamefont{and}
  \bibinfo{author}{\bibfnamefont{A.}~\bibnamefont{Rollett}},
  \emph{\bibinfo{title}{Mathematical Models}} (\bibinfo{publisher}{Tarquin
  Pub.}, \bibinfo{address}{Stradbroke, England}, \bibinfo{year}{1989}),
  \bibinfo{edition}{3rd} ed.

\bibitem[{\citenamefont{de~Gennes and Prost}(1995)}]{degennes.pg:1995.a}
\bibinfo{author}{\bibfnamefont{P.~G.} \bibnamefont{de~Gennes}}
  \bibnamefont{and} \bibinfo{author}{\bibfnamefont{J.}~\bibnamefont{Prost}},
  \emph{\bibinfo{title}{The Physics of Liquid Crystals}}
  (\bibinfo{publisher}{Clarendon Press}, \bibinfo{address}{Oxford},
  \bibinfo{year}{1995}), \bibinfo{edition}{second, paperback} ed.

\bibitem[{\citenamefont{Stephen and Straley}(1974)}]{stephen.mj:1974.a}
\bibinfo{author}{\bibfnamefont{M.}~\bibnamefont{Stephen}} \bibnamefont{and}
  \bibinfo{author}{\bibfnamefont{J.}~\bibnamefont{Straley}},
  \bibinfo{journal}{Rev. Mod. Phys.} \textbf{\bibinfo{volume}{46}},
  \bibinfo{pages}{617} (\bibinfo{year}{1974}).

\bibitem[{\citenamefont{de~Gennes}(1971)}]{degennes.pg:1971.a}
\bibinfo{author}{\bibfnamefont{P.~G.} \bibnamefont{de~Gennes}},
  \bibinfo{journal}{Mol Cryst Liquid Cryst} \textbf{\bibinfo{volume}{12}},
  \bibinfo{pages}{193} (\bibinfo{year}{1971}).

\bibitem[{\citenamefont{West et~al.}(2002)\citenamefont{West, Glushchenko,
  Liao, Reznikov, Andrienko, and Allen}}]{west.jl:2002.a}
\bibinfo{author}{\bibfnamefont{J.~L.} \bibnamefont{West}},
  \bibinfo{author}{\bibfnamefont{A.}~\bibnamefont{Glushchenko}},
  \bibinfo{author}{\bibfnamefont{G.}~\bibnamefont{Liao}},
  \bibinfo{author}{\bibfnamefont{Y.}~\bibnamefont{Reznikov}},
  \bibinfo{author}{\bibfnamefont{D.}~\bibnamefont{Andrienko}},
  \bibnamefont{and} \bibinfo{author}{\bibfnamefont{M.~P.} \bibnamefont{Allen}},
  \bibinfo{journal}{Phys. Rev. E} \textbf{\bibinfo{volume}{66}},
  \bibinfo{pages}{012702} (\bibinfo{year}{2002}).

\bibitem[{\citenamefont{Andrienko et~al.}(2004)\citenamefont{Andrienko,
  Tasinkevych, Patr{\'\i}cio, and {Telo da Gama}}}]{andrienko:2004.a}
\bibinfo{author}{\bibfnamefont{D.}~\bibnamefont{Andrienko}},
  \bibinfo{author}{\bibfnamefont{M.}~\bibnamefont{Tasinkevych}},
  \bibinfo{author}{\bibfnamefont{P.}~\bibnamefont{Patr{\'\i}cio}},
  \bibnamefont{and} \bibinfo{author}{\bibfnamefont{M.~M.} \bibnamefont{{Telo da
  Gama}}}, \bibinfo{journal}{Phys. Rev. E} \textbf{\bibinfo{volume}{69}},
  \bibinfo{pages}{021706} (\bibinfo{year}{2004}).

\bibitem[{\citenamefont{Binder}(1994)}]{binder:1994}
\bibinfo{author}{\bibfnamefont{K.}~\bibnamefont{Binder}},
  \bibinfo{journal}{Theories And Mechanism Of Phase Transitions, Heterophase
  Polymerizations, Homopolymerization, Addition Polymerization}
  \textbf{\bibinfo{volume}{112}}, \bibinfo{pages}{181} (\bibinfo{year}{1994}).

\bibitem[{\citenamefont{Flebbe et~al.}(1996)\citenamefont{Flebbe, D{\"u}nweg,
  and Binder}}]{flebbe.t:1996.a}
\bibinfo{author}{\bibfnamefont{T.}~\bibnamefont{Flebbe}},
  \bibinfo{author}{\bibfnamefont{B.}~\bibnamefont{D{\"u}nweg}},
  \bibnamefont{and} \bibinfo{author}{\bibfnamefont{K.}~\bibnamefont{Binder}},
  \bibinfo{journal}{J. Phys. II} \textbf{\bibinfo{volume}{6}},
  \bibinfo{pages}{667} (\bibinfo{year}{1996}).

\bibitem[{\citenamefont{Olsson et~al.}(2004)\citenamefont{Olsson, Linse, and
  Piculell}}]{olsson.m:2004}
\bibinfo{author}{\bibfnamefont{M.}~\bibnamefont{Olsson}},
  \bibinfo{author}{\bibfnamefont{P.}~\bibnamefont{Linse}}, \bibnamefont{and}
  \bibinfo{author}{\bibfnamefont{L.}~\bibnamefont{Piculell}},
  \bibinfo{journal}{Langmuir} \textbf{\bibinfo{volume}{20}},
  \bibinfo{pages}{1611} (\bibinfo{year}{2004}).

\bibitem[{\citenamefont{Boulbitch}(2002)}]{boulbitch.a:2002.a}
\bibinfo{author}{\bibfnamefont{A.}~\bibnamefont{Boulbitch}},
  \bibinfo{journal}{Europhysics Letters} \textbf{\bibinfo{volume}{59}},
  \bibinfo{pages}{910} (\bibinfo{year}{2002}).

\bibitem[{\citenamefont{Deserno and Bickel}(2003)}]{dezerno.m:2003.a}
\bibinfo{author}{\bibfnamefont{M.}~\bibnamefont{Deserno}} \bibnamefont{and}
  \bibinfo{author}{\bibfnamefont{T.}~\bibnamefont{Bickel}},
  \bibinfo{journal}{Europhysics Letters} \textbf{\bibinfo{volume}{62}},
  \bibinfo{pages}{767} (\bibinfo{year}{2003}).

\bibitem[{\citenamefont{Dietrich}(1988)}]{dietrich.s:1988.a}
\bibinfo{author}{\bibfnamefont{S.}~\bibnamefont{Dietrich}}, in
  \emph{\bibinfo{booktitle}{Phase Transitions and Critical Phenomena}}, edited
  by \bibinfo{editor}{\bibfnamefont{C.}~\bibnamefont{Domb}} \bibnamefont{and}
  \bibinfo{editor}{\bibfnamefont{J.}~\bibnamefont{Lebowitz}}
  (\bibinfo{publisher}{Academic Press}, \bibinfo{year}{1988}),
  vol.~\bibinfo{volume}{12}, p.~\bibinfo{pages}{1}.

\end{thebibliography}

\end{document}